\documentclass[12pt]{revtex4}
 \begin{document}
 \vskip -1in
 \centerline{\bf \Large A Few Recollections of Stephen}
 \centerline{\bf \large James Hartle}
 
 {\it On the evening after Stephen Hawking's funeral in Cambridge on March 31, 2018  a dinner for attendees who had come from far away was hosted by Paul Shellard, the Director of the Centre for Theoretical Cosmology. Paul asked me to speak for five minutes on my recollections of Stephen. The following is an slightly edited copy of my speaking text.} 
 \vskip .1in
 
 \large
Paul asked me to speak for a few minutes on my recollections of Stephen. That's impossible. I have too  many of them. I picked only three  to illustrate various aspects of Stephen's persona \footnote{  For more on the author's personal experiences working with Stephen see ``Working with Stephen'', arXiv:1711.09071 and  the authorÕ's contribution to the March 14 article in {\sl Physics Today} in which Stephen Hawking is remembered
by his colleagues. https://goo.gl/dRgf3q), arXiv:1803.09197}. 

In 1971-1972 I made a long visit to Fred Hoyle's Institute of Theoretical Astronomy (as it was known then). Pulsars, quasars, and the cosmic background radiation had been recently discovered.  I  used a Sloan Fellowship to go to Cambridge and retrain in the emerging field of  relativistic astrophysics.  There I  met  Martin Rees, Brandon Carter, Paul Davis and had my first scientific interactions with Stephen Hawking.  Right away we seemed to be on the same wavelength. There we wrote what was to be the first of eleven joint papers over the years. Many more visits to Cambridge were to follow. 

These recollections come mostly from a time when Stephen could still walk a little with assistance, when his wheelchair was a ordinary one, when he could still speak, and when he could get by during the day  with only the help of those of us around him. 
 
{\it Confidence:}  Stephen had a powerful and accurate physical intuition and a remarkable level of confidence and certainty.  As an example, in those happier days,  my wife and I packed Stephen and chair into the back of our VW beetle to get to the airport for a flight to Copenhagen where the big GR conference was to be held.  We were driving, but it was Stephen who was giving the directions from the back seat, sure that he knew the fastest way to  Heathrow. It seemed like all left turns  to me but we made it.  

{\it Breaking Rules:} Stephen was not afraid of breaking rules in science and also in  life.  For instance, black holes are not black.  At that time he was still driving in a kind of three wheel vehicle for the disabled.  It was a long walk for me from the Madingley road to my rented flat in Cambridge. Stephen would give me lifts into town with  me perched in the back of this contraption.   Certainly illegal, possibly dangerous, but definitely Stephen. 

{\it Friendliness: } Stephen and Jane were very welcoming and my wife and I were often over to Little St. Mary's or
 5 West Road. On one occasion they even staged an American barbecue to make us feel at home. There was a little pile of charcoal, some ground meat.  And there were hot cross buns to take the place of hamburger buns --- something not known here then. Stephen and Jane overcoming all obstacles. 

{\it Science:} My fondest recollection is about science.  Stephen once said to me:  ``I think the no- boundary wave function is the best thing that either of us has done.''  I don't think that  I will ever receive a more treasured accolade. 

It  has been said  that ``Man is not eternal save in the consequences of his work and the memories of his friends.'' Knowing  the impact of Stephen's work, and seeing the many people who have come to Cambridge on this occasion, I think that Stephen is doing well on both counts!

As scientists, Stephen will be ever in our minds because he has given us so much to think about. But for those of us who had the privilege of working with him he will also be --- ever in our hearts.  

\end{document}